\documentclass{aa} 
\usepackage{graphicx} 
\usepackage{txfonts}  
\usepackage{natbib} 
\begin{document} 
\title{The population of barred galaxies in the local universe} 

\subtitle{I. Detection and characterisation of bars}  

\author{J. A. L. Aguerri\inst{1} 
 \and 
 J. M\'endez-Abreu \inst{2,3,4} 
 \and 
 E. M. Corsini \inst{4}} 

\offprints{J. A. L. Aguerri} 
 
\institute{Instituto de Astrof\'\i sica de 
       Canarias, C/ V\'ia L\'actea s/n, E-38200 La Laguna, Spain\\ 
         \email{jalfonso@iac.es}  
  \and INAF-Osservatorio Astronomico di Padova, vicolo 
       dell'Osservatorio~5, I-35122 Padova, Italy\\ 
         \email{jairo.mendez@oapd.inaf.it}  
  \and Universidad de La Laguna, Av. Astrof\'isico Francisco 
       S\'anchez s/n, E-38206 La Laguna, Spain  
  \and Dipartimento di Astronomia, Universit\`a di Padova, vicolo 
       dell'Osservatorio~3, I-35122 Padova, Italy\\  
         \email{enricomaria.corsini@unipd.it}  
         } 
\date{Received ; accepted } 
 
\authorrunning{Aguerri,  M\'endez-Abreu, and Corsini} 
\titlerunning{Detection and characterisation bars in the local universe}

\abstract 
{Bars are very common in the centre of the disc galaxies, and they drive the
  evolution of their structure. The state-of-the-art imaging and redshift
  surveys of galaxies allow us to study the relationships between the
  properties of the bars and those of their hosts in statistically
  significant samples.}
{A volume-limited sample of 2106 disc galaxies was studied to derive
  the bar fraction, length, and strength as a function of the
  morphology, size, local galaxy density, light concentration, and
  colour of the host galaxy. The sample galaxies were selected to not be
  strongly disturbed/interacting.}
{The bar and galaxy properties were obtained by analysing the $r-$band
  images of the sample galaxies available in the Sloan Digital Sky
  Survey Data Release 5.}
{The bars were detected using the ellipse fitting method and Fourier
  analysis method. They were tested and calibrated with extensive
  simulations on artificial images. The ellipse fitting method was
  found to be more efficient in detecting bars in spiral galaxies. The
  fraction of barred galaxies turned out to be $45\%$. A bar was found in
  $29\%$ of the lenticular galaxies, in $55\%$ and $54\%$ of the
  early- and late-type spirals, respectively.
The bar length (normalised by the galaxy size) of  late-type spirals is
shorter than  in early-type or lenticular ones.  A correlation between
the bar  length and galaxy size  was found with longer  bars hosted by
larger galaxies.
The bars of the lenticular  galaxies are weaker than those in spirals.
Moreover, the unimodal distribution of  the bar strength found for all
the galaxy types argues against  a quick transition between the barred
and unbarred statues.
There is no difference between  the local galaxy density of barred and
unbarred  galaxies. Besides, neither  the length  nor strength  of the
bars   are  correlated   with  the   local  density   of   the  galaxy
neighbourhoods.
In contrast, a statistical  significant difference between the central
light  concentration and colour  of barred  and unbarred  galaxies was
found.   Bars  are  mostly  located  in less  concentrated  and  bluer
galaxies.}
{These  results indicate  that  the properties  of  bars are  strongly
related to  those of their host  galaxies, but do not  depend on the
local environment.}


\maketitle 
%
 
\section{Introduction} 
 
Strong bars are observed in optical  images of roughly half of all the
nearby disc galaxies \citep[][]{marinova07, reese07, barazza08}.  This
fraction rises to about  $70\%$ when near-infrared images are analysed
\citep[][]{knapen00, eskridge00, menendezdelmestre07}.
The   presence  of   a  bar   can  be   found  by   visual  inspection
\citep[e.g.,][hereafter RC3]{devaucouleurs91},  by analysing the shape
and  orientation of  the  galaxy isophotes  \citep[e.g.,][]{wozniak95,
laine02, marinova07,  barazza08}, or by studying the  Fourier modes of
the light distribution \citep[e.g.,][]{ohta90, elmegreen85, aguerri98,
aguerri00a}.
Therefore, bars  are a common feature  in the central  regions of disc
galaxies  of the  local  universe.  But,  this  is also  true at  high
redshift. In fact, the bar fraction apparently remains constant out to
$z\approx  1$  \citep[][]{jogee04,  elmegreen04, barazza08},  although
there are some claims that this is not the case \citep{sheth08}.

The presence of  bars in the centre of  lenticular and spiral galaxies
make  them ideal  probes of  the dynamics  of the  central  regions of
discs.   In  fact, bars  are  efficient  agents  of angular  momentum,
energy, and  mass redistribution. They  act on both luminous  and dark
matter   components  \citep{weinberg85,   debattista98,  debattista00,
athanassoula03}  driving   the  evolution  of   galaxy  structure  and
morphology.  In  particular, the amount of  angular momentum exchanged
is related  to specific  properties of the  galaxies, such as  the bar
mass,     halo    density,     and     halo    velocity     dispersion
\citep{athanassoula03, sellwood06, sellwooddebattista06}
Moreover,  they funnel  material  towards the  galaxy centre  building
bulge-like  structures  \citep[e.g.,][]{kormendykennicutt04},  nuclear
star-forming   rings   \citep[e.g.,][]{buta03},   and   nuclear   bars
\citep[e.g.,][]{erwin04},   and  feeding   the   central  black   hole
\citep[e.g.,][]{shlosman00}.

    Bars play an important role in bulge formation. Major mergers or
    monolithic collapse are the classical theories for bulge formation
    in disc galaxies \citep[e.g.,][]{eggen62, kormendykennicutt04}.  But bulges
    can be built via minor mergers \citep[][]{aguerri01a,
      elichemoral06} or secular evolution processes produced by bars.
    In fact, according with the results of N-body simulations, the
    inner parts of a bar inflate after a few bar rotations because of
    large-scale violent bending instabilities and settle with an
    increased thickness and vertical velocity dispersion
    \citep[e.g.,][]{combes81, combes90, raha91, athanassoula03,
      debattista04, athanassoula05, martinezvalpuesta06}.  This leads
    to the establishment of the connection between the bar-buckling
    mechanism and the formation of boxy/peanut bulges
    \citep[][]{bureau99a, bureau99b, chung04}.  The buckling
    instability does not destroy the bar and forms a central stellar
    condensation reminiscent of the bulges of late-type spirals
    \citep[][]{debattista04, athanassoula05}, in agreement with early
    findings by \citet[][]{hohl71}.  Observational evidences of
    secular bulge formation includes the near-exponential surface
    brightness profiles of some bulges \citep[][]{andredakis95,
      courteau96, dejong96, carollo01, prieto01, macarthur03,
      aguerri05, mendezabreu08a, fisher08}, a correlation between
    bulge and disc scale length \citep[][]{macarthur03, aguerri05,
      mendezabreu08a}, the similar colours of bulges and inner discs
    \citep[][]{peletier96, courteau96, carollo07}, substantial
    rotation \citep[][]{kormendykennicutt04}, and the presence of
    B/P-shaped bulges in $\approx 45\%$ of edge-on galaxies
    \citep[][]{lutticke00}.  Recently, the connection between
    B/P-shaped bulges and bars has also been confirmed in face-on
    barred galaxies \citep[][]{mendezabreu08c}.

The most  important parameters of  bars are the length,  strength, and
pattern  speed. Their evolution  depends on  the effectiveness  of the
angular momentum exchange between luminous and dark matter.
Different methods have been proposed to measure  bar properties.

The  bar length  can be  obtained by  eye estimates  on  galaxy images
\citep[][]{kormendy79, martin95}, locating  the maximum ellipticity of
the  galaxy   isophotes  \citep[][]{wozniak95,  laine02,  marinova07},
looking   for    variations   of   the    isophotal   position   angle
\citep[][]{sheth03,  erwin05} or  of the  phase angle  of  the Fourier
modes   of  the   galaxy   light  distribution   \citep[][]{quillen94,
aguerri03}, analysing the bar-interbar contrast \citep[][]{aguerri00a,
aguerri03}, or by photometric  decomposition of the surface brightness
distribution \citep[][]{prieto97, aguerri05, laurikainen05}.
The previous techniques reported that the typical bar length is about
3-4 kpc,  and is correlated with  the disc scale,  suggesting that the
two  components   are  affecting  each   other  \citep[][]{aguerri05, marinova07,
laurikainen07}.

The  bar  strength  can  be  derived  by  measuring  the  bar  torques
\citep[][]{buta01},   isophotal   ellipticity   \citep[][]{martinet97,
aguerri99, whyte02,  marinova07}, the  maximum amplitude of  the $m=2$
Fourier mode \citep[][]{athanassoula02, laurikainen05}, or integrating
the  $m=2$   Fourier  mode   in  the  bar   region  \citep[][]{ohta90,
aguerri00a}.
The   bar    strength   is   almost   constant    with   Hubble   type
\citep[][]{marinova07}, but lenticular  galaxies host weaker bars than
spirals \citep[][]{laurikainen07}.

The pattern speed  of bars can be indirectly  estimated by identifying
rings    with    the    location    of   the    Lindblad    resonances
\citep[e.g.,][]{vegabeltran97,   jeong07},   matching   the   observed
velocity and  density fields  with numerical models  of the  gas flows
\citep[e.g.,][]{lindblad96, aguerri01b, weiner01, rautiainen08}, 
analysing the offset and shape of the dust lanes, which trace the
location of shocks in the gas flows \citep[e.g.,][]{athanassoula92},
looking for colour changes
\citep[][]{aguerri00a}    and   minima    in   the    star   formation
\citep[][]{cepa90}  outside  the  bar   region,  or  by  adopting  the
Tremaine-Weinberg  method \citep[see][for a  review]{corsini07b}.  The
last  is  a  model-independent   way  to  measure the   pattern  speed
\citep[][]{tremaine84},  which  was  successfully  applied  to  single
 \citep[][]{merrifield95,debattista02, aguerri03}, double
\citep[][]{corsini03} and dwarf barred galaxies \citep[][]{corsini07}
too.
Observed pattern speeds imply  that barred galaxies host maximal discs
\citep{debattista98,  debattista00}, since bars  in dense  dark matter
halos are rapidly decelerated by dynamical friction \citep{weinberg85,
sellwood06, sellwooddebattista06}.

Nowadays,  the large  galaxy surveys,  such as  the Sloan  Digital Sky
Survey  \citep[][hereafter SDSS]{york00}  allow  us to  study the  bar
properties  in samples  of thousands  of galaxies,  obtaining  for the
first   time  statistically   significant  results.   By   studying  a
volume-limited  sample of  $\sim$3000 galaxies  in the  local universe
extracted from  the SDSS,  we plan to  address three main  issues: the
first is to assess, by means of extensive tests on simulated galaxies,
the  advantages and  drawbacks of  the  two more  common methods  used
detecting  bars: the ellipse  fitting and  the Fourier  analysis.  The
second  is  to  investigate   the  possible  differences  of  the  bar
fraction and bar properties  (length  and strength)  with  the morphological
type, ranging from S0 to late-type spirals. The third is to understand
how the properties  of the host galaxies affects  the formation of the
bar and its properties.

The paper is  organised as follows. The galaxy  sample is presented in
Sect.~\ref{sec:sample}; the  methods we adopted to detect  the bars in
the sample galaxies are explained in Sect.~\ref{sec:methods}; they are
tested  using   artificial  galaxy  images   in  Sect.~\ref{sec:test};
the fraction  of bars are given  in Sect.~\ref{sec:fraction}; the
bars   properties    (length,   and   strength)    are   reported   in
Sect.~\ref{sec:bar_properties} and compared  to galaxy properties in
Sect.~\ref{sec:galaxy_properties};    conclusions    are   given    in
Sect.~\ref{sec:conclusions}.    Throughout   this   paper  we   assume
$H_{0}$=100 km s$^{-1}$ Mpc$^{-1}$.

\section{Sample selection and data reduction} 
\label{sec:sample} 
 
The sample  galaxies were selected  in the spectroscopic  catalogue of
the SDSS Data Release 5 \citep[SDSS-DR5,][]{adelman07}.
From the  $\sim$675,000 galaxies available  in the catalogue,  we took
all the galaxies in  the redshift range $0.01 < z <  0.04$ and down to
an   absolute   magnitude    $M_{r}   <   -20$   \citep[$\approx
M^{*}_{r}$,][]{blanton05}.   This represents a  volume-limited sample,
because  the apparent  magnitude of  a galaxy  with $M_{r}  =  -20$ at
$z=0.04$ ($m_{r} \sim 15.5$)  is within the completeness limit ($m_{r}
= 17.77$) of the SDSS-DR5 spectroscopic catalogue.

In order to deal with  projection effects, we restricted our sample to
galaxies  with  $b/a>0.5$,  $a$  and  $b$  being  the  semi-major  and
semi-minor axis  lengths of the  galaxies. For disc galaxies,  this is
equivalent  to say  that  we have  selected  objects with  inclination
$i<60^{\circ}$. Although  the cut in  the observed axial ratio  of the
sample galaxies introduces  a bias in the selection  of the elliptical
ones,  this will  not affect  the results  of the  paper since  we are
interested only in the properties of disc galaxies.

Then, we rejected all galaxies with close neighbours. To this
aim, we excluded all the galaxies with a companion which was closer
than $2 \times r_{90}$, where $r_{90}$ is defined as the radius which
contains $90\%$ of the total galaxy light. In addition, the companion
must be within $\pm$ 3 magnitudes with respect to the magnitude
of the target galaxy to be excluded. In this way, galaxies with faint
companions or possibly contaminated by faint
foreground/background objects are not discarded in our study.  The
resulting sample consisted of 3060 galaxies.

According to \citet{marinova07} the bar length of barred galaxies in
the local universe ranges between about 0.5 and 5 kpc, with a
    mean value $\sim$3.5 kpc. At $z=0.04$ the shortest bar length
    ($r_{\rm bar} = 0.5$ kpc) projects onto 2.17 pixel ($0\farcs86$)
    in the SDSS images, which have a scale of $0\farcs3946$ arcsec
    pixel$^{-1}$.  The PSF of the SDSS images can be modelled
assuming a Moffat function \citep[see e.g.][]{trujillo01}.  We fit a
bidimensional Moffat function of the form

\begin{equation}
{\rm PSF}(r)=\frac{\beta -1}{\pi \alpha^2}\left(1+\frac{r}{\alpha}^2\right)^{-\beta},
\end{equation}
%
to several stars in each galaxy field obtaining a typical FWHM and
$\beta$ parameter of $1\farcs09$ (2.77 pixel) and 3.05, respectively.
This means that the bars with a length of 0.5 kpc are not resolved. According to the tests on artificial galaxies we performed (see
Sect. 4), the smallest bars that we are able to recover in the SDSS
images have a length of $\sim 9$ pixel. This corresponds to $\sim 0.5$
kpc at $z=0.01$ and $\sim 2$ kpc at $z=0.04$. Therefore, a value of 2
kpc is a more reliable limit on the actual resolution of the bar
length throughout the range of distances covered by our sample
galaxies.

\begin{figure} 
   \centering 
   \includegraphics[width=9cm]{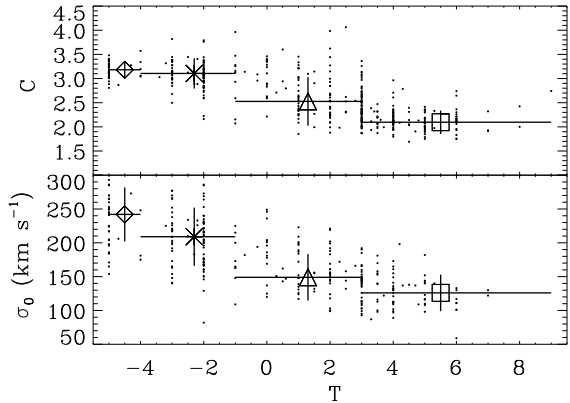} 
      \caption{Values of the light concentration $C$ (top panel) and
        central velocity dispersion $\sigma_{0}$ (bottom panel) as a
        function of the morphological parameter $T$. Only the sample
        galaxies in RC3 are plotted. Median values of $C$ and
        $\sigma_{0}$ for ellipticals (diamond), lenticulars
        (asterisk), early-type spirals (triangle), and late-type
        spirals (square) are shown. }
    \label{fig:morphology_rc3} 
    \end{figure}

The visual morphological classification given by the RC3 was available
only for a subsample of 612 galaxies.
Automatic   galaxy  morphological   classifications   divide  galaxies
according to  some photometric  observables. In particular,  the light
concentration   is   strongly   correlated   with  the   Hubble   type
\cite[e.g.,][]{abraham96, conselice00,  conselice03}.  In fact,  it is
greater in  early- than  in late-type galaxies.   We defined  the light
concentration  as $C=r_{90}/r_{50}$, where  $r_{50}$ and  $r_{90}$ are
the  radii enclosing  $50\%$ and  $90\%$  of the  total galaxy  light,
respectively.  These radii are available  in the SDSS database for all
objects of our sample.
We calculated  the median values of  $C$ for the  ellipticals ($T \leq
-4$), lenticulars ($-3 \leq T \leq -1$), early-type spirals ($0 \leq T
\leq 3$),  and late-type  spirals ($T \geq  4$) of the  RC3 subsample.
They  are   given  in  Table~\ref{tab:morphology_rc3}   and  shown  in
Fig.~\ref{fig:morphology_rc3}. Although the  dispersion of the data is
large,  the median  value of  the light  concentration  decreases with
increasing Hubble type.

If the light  traces mass, then we expect a  relation between the mass
concentration   and  morphological   type  too.    The   central  mass
concentration  of a  galaxy  can  be traced  by  the central  velocity
dispersion $\sigma_{0}$.   This was available  in the SDSS for  298 of
the sample galaxies listed in RC3.
We calculated the median values of $\sigma_0$ for the different Hubble
types, and they are  given in Table~\ref{tab:morphology_rc3} and shown
in  Fig.~\ref{fig:morphology_rc3}.   The  median  velocity  dispersion
decreases with increasing Hubble type, too.

\begin{figure} 
   \centering 
   \includegraphics[width=9cm]{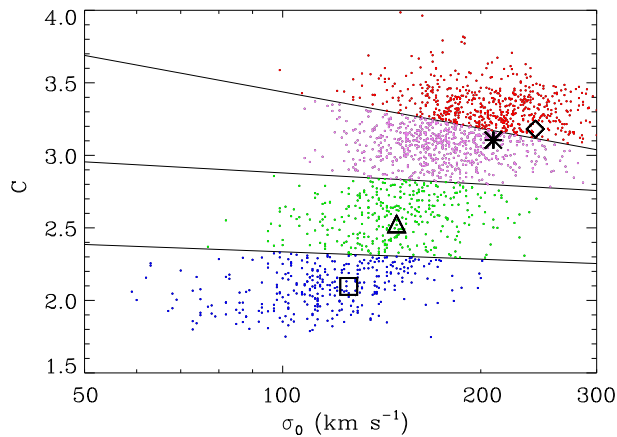} 
      \caption{Values of the light concentration $C$ and central
        velocity dispersion $\sigma_{0}$ for the sample galaxies with
        $\sigma_{0}$ available in SDSS-DR5.  They are colour coded
        according to our morphological classification: the red,
          magenta, green, and blue dots correspond to the
        ellipticals, lenticulars, early-type and late-type spirals,
        respectively.  Median values for the ellipticals (diamond),
        lenticulars (asterisk), early-type spirals (triangle), and
        late-type spirals (square) plotted in Fig.~1 are also shown.
      The full lines represent the locus of equal distance between different galaxy morphological types.}
   \label{fig:csigma} 
   \end{figure}

Fig. \ref{fig:csigma} shows the  relation between $C$ and $\sigma_{0}$
for  those galaxies  in our  sample  with both  the photometric  and
kinematic information.
Each galaxy  can be assigned to  a morphological bin  according to its
values of  $C$ and  $\sigma_{0}$. They were  assigned to the  bin with
closest   median  values  listed   in  Table~\ref{tab:morphology_rc3} by minimizing the following
equation:

\begin{equation}
d=\sqrt{(C-C')^{2}+(log(\sigma_{0})-log(\sigma_{0}'))^{2}},
\end{equation} 
where $C'$ and $\sigma_{0}'$ are the median values of $C$ and
$\sigma_{0}$ reported in Table~\ref{tab:morphology_rc3}.

Galaxies, for  which only the light concentration  was available, were
assigned to the morphological  bin corresponding to the closest median
value of $C$.

We found that $26\%$, $29\%$, $20\%$, and $25\%$ of the selected
galaxies turned to be ellipticals, lenticulars, early-type and
late-type spirals, respectively. In this work, we focused on the 2166
disc galaxies which include the lenticulars and the early-type
  and late-type spirals. This represents our final sample.  Among the
disc galaxies $39\%$, $28\%$, and $33\%$ were classified as
lenticulars, early-type and late-type spirals, respectively.
Table~\ref{tab:missclassify} shows the comparison between the RC3
morphological classification and our automatic classification for the
sample galaxies listed in RC3.  It lists the fraction of elliptical,
S0, S0/a-Sb and Sbc-Sm galaxies as classified by RC3 among the
galaxies we classified as lenticular, early- and late-type spirals.
It is worth noticing that for all the disc galaxies more than
    50$\%$ of the objects are assigned to the same morphological bin
    by both the RC3 and our automatic classification. The better
    agreement is shown by the early-type spiral
    galaxies. Nevertheless, the contamination between the different
    classes could be strong.

The  $r-$band  image  of  each  galaxy was  retrieved  from  the  SDSS
archive. All  the images  were bias subtracted,  flat-field corrected,
and sky subtracted according to the associated calibration information
stored in the Data Archive Server (DAS).
%
\begin{table} 
\caption{Median values of the light concentration and central
velocity dispersion for the different galaxy types.}
\label{tab:morphology_rc3} 
\centering 
\begin{tabular}{c c c}     
\hline\hline 
\noalign{\smallskip}
Galaxy Type & $C$ & $\sigma_{0}$ \\
            &     & (km s$^{-1}$) \\ 
\noalign{\smallskip}
\hline 
\noalign{\smallskip}
E       & 3.18$\pm$0.15 & 242$\pm$40 \\  
S0      & 3.10$\pm$0.31 & 209$\pm$43 \\ 
S0/a-Sb & 2.53$\pm$0.50 & 149$\pm$34 \\ 
Sbc-Sm  & 2.10$\pm$0.24 & 126$\pm$27 \\ 
\noalign{\smallskip}
\hline 
\end{tabular} 
\end{table} 
\begin{table} 
\caption{Comparison between our and RC3 classification.} 
\label{tab:missclassify} 
\centering 
\begin{tabular}{l c c c}     
\hline\hline 
   & S0 (our) & Early-type spirals (our) & Late-type spirals (our) \\ 
\hline 
E          (RC3) & 0.16     &  0.01        &  0.01   \\  
S0         (RC3) & 0.53     &  0.05        &  0.03   \\ 
S0/a-Sb (RC3) & 0.26     &  0.67        &  0.37   \\ 
Sbc-Sm  (RC3) & 0.05     &  0.27        &  0.59   \\ 
\hline 
\end{tabular} 
\end{table} 

\section{Methods for detecting and analysing bars} 
\label{sec:methods}
 
Three main methods have been proposed for detecting bars and analysing
their properties \citep[see  for a review][]{erwin05, michel06}.  They
are   based   on   the   ellipse   fit   of   the   galaxy   isophotes
\cite[][]{wozniak95, knapen00,  laine02, sheth03, elmegreen04,jogee04,
marinova07, barazza08},  Fourier analysis of  the azimuthal luminosity
profile    \cite[][]{elmegreen85,    ohta90,    aguerri00a,    buta06,
laurikainen07},  and decomposition  of  the galaxy  surface-brightness
distribution \cite[][]{prieto01, aguerri05, laurikainen05}.

In the present work, we have developed a fully automatic method
  for classifying barred and unbarred galaxies using the ellipse
  fitting (Sect.  \ref{sec:ellipse}) and Fourier analysis (Sect.
  \ref{sec:fourier}). This automatic procedure has several advantages
  with respect to the user-dependent visual classifications. In fact,
  it is reproducible, it can be implemented and applied to large data
  sets. We first tested it by means of extensive simulations on a
  large set of artificial disc galaxies (Sect. \ref{sec:test}) and
  then we applied it to the images of the sample galaxies.
In a  forthcoming paper we  will obtain the structural  parameters for
the bulge,  disc, and  bar of  all the sample  galaxies by  applying a
photometric  decomposition  of  their surface-brightness  distribution
based on the technique developed by \citet{mendezabreu08b}.

\subsection{Ellipse fitting method} 
\label{sec:ellipse} 
 
This method is based on the fit of the galaxy isophotes by ellipses. 

As a first step, each image was cleaned of field stars and
    galaxies.  This was done by rotating the image by 180$^{\circ}$
    with respect to the galaxy centre. Then, we subtract the rotated
    frame to the original one. The residual image was sigma-clipped to
    identify all the pixels with a number of counts lower than
    $1\sigma$, where $\sigma$ is the rms of the image background after
    sky subtraction and calculated in regions free of sources and far
    from the galaxy to avoid contamination. The value of the deviant
    pixels was set to zero. Finally, the clipped image was subtracted
    to the original one to get the cleaned and symmetrized image to
    be used in the analysis.

The ellipses were fitted to the isophotes of the cleaned and
symmetrized images of the 2166 disc galaxies using the IRAF\footnote{IRAF
  is distributed by NOAO, which is operated by AURA Inc., under
  contract with the National Science Foundation.} task ELLIPSE
\citep[][]{jedrzejewski87}.
In order  to get a  good fit  at all radii  out to an  intensity level
corresponding to the background rms, we implemented the fitting method
described by \citet{jogee04} and \citet{marinova07}.
This is an iterative wrapped procedure, which runs the ellipse fitting
several times changing the trial values at each fit iteration.
At each fixed semi-major axis length, the coordinates of the
    centre of the fitting ellipse were kept fixed and corresponded to
    those of the galaxy centre. This was identified with the position
    of the intensity peak. The trial values for the ellipticity
    $\epsilon$ and position angle PA were randomly chosen between 0 and 1 and between
    $-90^\circ$ and $90^\circ$, respectively. The fitting procedure
    stopped when either the convergence was reached or after 100
    iterations.  The ellipse fit failed for a small fraction (60/2166) of the sample galaxies.  For all the galaxies with
    properly fitted isophotes, we obtained the radial profiles of
$\epsilon$ and PA of the fitted ellipses.

The ellipticity radial profile in a bright and inclined unbarred
    galaxy usually shows a global increase from low values in the
centre to a constant value at large radii.  At large radii the PA
radial profile is constant too.  The constant values of $\epsilon$ and
PA (on large radial scales) are related to the inclination and
orientation of the line of nodes of the galactic disc.  On the
contrary, barred galaxies are characterised by the presence of a local
maximum in the ellipticity radial profile and constant PA in the bar
region \citep[see e.g.,][]{wozniak95, aguerri00b}.  This is due to the
shape and orientation of the stellar orbits of the bar
\citep[see][]{contopoulos89, athanassoula92}.

This allowed us  to identify bars by analysing  the radial profiles of
$\epsilon$ and PA.  We considered that  a galaxy hosts a bar when: (1)
the ellipticity  radial profile shows a  significant increase followed
by a significant  decrease ($\Delta \epsilon \geq 0.08$),  and (2) the
PA of  the fitted ellipses is  roughly constant within  the bar region
($\rm  \Delta PA  \leq 20^\circ$)\footnote{We  have considered  that a
global maximum in the ellipticity profile is produced by a bar when it
is located at more than 3.5 pixels from the galaxy centre. This radial
distance correspond to more than 3 times the FWHM/2 of the images.}.

The  values adopted  for  $\Delta\epsilon$ and  $\rm  \Delta PA$  were
determined  by  applying  the   method  to  artificial  galaxies  (see
Sect. \ref{sec:ellipse_test}).
The bar length  was derived as the radius  $r_{\rm bar}^{\epsilon}$ at
which the  maximum ellipticity was  reached \citep[e.g.,][]{wozniak95,
laine02, marinova07} or as the  radius $r_{\rm bar}^{\rm PA}$ at which
the PA changes by $5^\circ$ with respect to the value corresponding to
the  maximum ellipticity  \citep[e.g,][]{wozniak95,  sheth03, erwin03,
erwin05, michel06}.

This method has been already successfully applied by different authors
both  in the  optical and  near-infrared wavebands  to detect  bars in
galaxies  at low  \citep[][]{knapen00, laine02,  marinova07}  and high
redshift \citep[][]{jogee04, elmegreen04, sheth08}.

\subsection{Fourier analysis method} 
\label{sec:fourier} 

An alternative way  to detect and characterise bars  is with a Fourier
analysis  of  the   azimuthal  luminosity  profile  \citep[][]{ohta90,
elmegreen85, aguerri00a}. 
The method starts by deprojecting the image of each sample galaxy by a
flux-conserving stretch along the minor axis by the factor $1/\cos i$,
where $i$ is the galaxy inclination.  Therefore, we needed a good
estimation of the inclination and major-axis PA of the galaxy. They
were derived from the ellipticity and PA of the ellipses fitted to the
five outermost isophotes, where the total luminosity is
  dominated by the disc contribution. For each galaxy, the intensity
  of the outermost isophote corresponds to a value of 1$\sigma$, where
  $\sigma$ is the rms of the sky-subtracted background of the image.

The deprojected luminosity  profile, $I(R,\phi)$, where $(R,\phi)$ are
the  polar coordinates  in  the  galaxy frame,  is  decomposed into  a
Fourier series
%
\begin{equation}
I(R,\phi)=\frac{A_{0}(R)}{2}+ \sum_{m=1}^\infty \left(A_{m}(R) 
\cos(m\phi)+B_{m}(R)\sin(m\phi)\right) ,
\label{eq:fourier_intensity}
\end{equation}
%
where the coefficients are defined by
%
\begin{equation}
A_{m}(R)=\frac{1}{\pi}\int^{2\pi}_{0}I(R,\phi)\cos(m\phi)d\phi,
\label{eq:fourier_coef_a}
\end{equation}
%
and
%
\begin{equation}
B_{m}(R)=\frac{1}{\pi}\int^{2\pi}_{0}I(R,\phi)\sin(m\phi)d\phi.
\label{eq:fourier_coef_b}
\end{equation}

The Fourier amplitude of the $m$-th component is defined as
 
\begin{equation} 
I_{m}(R) = \left\{ 
\begin{array}{ll}
  A_{0}(R)/2                       & \mbox{$m = 0$},  \\
  \sqrt{A_{m}^{2}(R)+B_{m}^{2}(R)} & \mbox{$m \neq 0$}. 
\end{array} 
\right.
\label{eq:fourier_amplitude}
\end{equation} 

The  even  ($m=2, 4,  6,  ...$)  relative  Fourier amplitudes  $I_{\rm
m}/I_{0}$ of galaxies with bars are  large, while the odd ($m=1, 3, 5,
...$) ones are small. In particular,  the bar is evidenced by a strong
$m=2$  component.   Similarly  to  the  ellipse   fitting  method,  we
considered  that a galaxy  hosts a  bar when:  (1) the  $m=2$ relative
Fourier  component  shows  a local  maximum  ($\Delta(I_{2}/I_{0})\geq
0.2$), and (2) the phase angle of the $m=2$ mode $\phi_{2}$ is roughly
constant within the bar  region ($\Delta \phi_{2} \leq 20^\circ$). The
values  adopted for $\Delta(I_{2}/I_{0})$  and $\Delta  \phi_{2}$ were
also  determined by applying  the method  to artificial  galaxies (see
Sect. \ref{sec:fourier_test}).

The bar  length $r_{\rm bar}^{\rm  Fourier}$ was calculated  using the
bar/interbar  intensity  ratio  as  in  \citet{aguerri00a}.   The  bar
intensity,  $I_{\rm b}$, is  defined as  the sum  of the  even Fourier
components, $I_{0}+I_{2}+I_{4}+I_{6}$,  while the inter-bar intensity,
$I_{\rm  ib}$, is  given  by $I_{0}-I_{2}+I_{4}-I_{6}$  \citep{ohta90,
elmegreen85, aguerri00a}.  \citet{ohta90}  arbitrarily defined the bar
length  as the  outer radius  for which  $I_{\rm b}/I_{\rm  ib}  = 2$.
However, \citet{aguerri00a} pointed out  that a fixed value of $I_{\rm
b}/I_{\rm ib}$ cannot account for the wide variety of bar luminosities
present in galaxies.  Instead, they defined the bar length as the FWHM
of  the curve of  $I_{\rm b}/I_{rm  ib}$. This  method was  applied by
\citet{athanassoula02} to  analytic models demonstrating  its accuracy
in measuring the bar length.

\section{Test on artificial galaxies} 
\label{sec:test} 

\subsection{Structural parameters of the artificial galaxies} 
\label{sec:parameters} 

Extensive simulations on a large  set of artificial disc galaxies were
carried  out to  test  the  reliability and  accuracy  of the  ellipse
fitting    (Sect.     \ref{sec:ellipse})    and    Fourier    analysis
(Sect. \ref{sec:fourier}) in detecting  bars. Moreover, they were used
to  fine tune the  free parameters  of the  two methods,  i.e., $\Delta
\epsilon$ and  $\Delta {\rm PA}$  in the ellipse fitting,  and $\Delta
(I_2/I_0)$ and $\Delta \phi_2$ in the Fourier analysis.
The  surface-brightness distribution  of the  artificial  galaxies was
assumed  to  be the  sum  of  the  contributions of  three  structural
components:  a  bulge, a  disc,  and  a bar  \citep[e.g.,][]{prieto01,
aguerri03,    aguerri05,   laurikainen05}.    The   surface-brightness
distribution  of each  individual component  was assumed  to  follow a
parametric law,  which has to  be strictly considered as  an empirical
fitting function.

The  S\'ersic  law \citep[][]{sersic68}  was  assumed  for the  radial
surface-brightness profile of the bulge
%
\begin{eqnarray}
I_{\rm bulge}(r)\;=\;I_{\rm 0,bulge}\;10^{-b_{n}\left(r/r_{\rm
e}\right)^{1/n}},
\label{eq:bulge}
\end{eqnarray}
%
where $r_{\rm e}$, $I_{\rm 0, bulge}$, and $n$ are the effective (or
half-light) radius, the central surface brightness, and the shape
parameter describing the curvature of the profile, respectively. The
value of $b_n$ is coupled to $n$ so that half of the total flux is
always within $r_{\rm e}$ and can be approximated as
$b_{n}=0.868n-0.142$ \citep{caon93}.  The total luminosity of the
bulge is given by
%
\begin{equation} 
L_{\rm bulge} = 2 \pi I_{\rm 0,bulge} (1-\epsilon_{\rm bulge}) n 
r_{\rm e}^{2}\frac{\Gamma(2n)}{b_{n}^{2n}}, 
\label{eq:total_bulge}
\end{equation} 
%
where $\epsilon_{\rm bulge}$ is  the observed ellipticity of the bulge
and $\Gamma$ is the Euler gamma function.

The exponential law \citep[][]{freeman70}  was assumed to describe the
radial surface-brightness profile of the disc
%
\begin{equation}
I_{\rm disc}(r)=I_{\rm 0,disc} e^{-r/h},
\label{eq:disc}
\end{equation}
\noindent 
%
where  $h$ and  $I_{\rm  0,disc}$  are the  scale  length and  central
surface brightness of the  disc, respectively. The total luminosity of
the disc is given by
%
\begin{equation} 
L_{\rm disc} = 2 \pi I_{\rm 0,disc}(1-\epsilon_{\rm disc})h^{2},
\label{eq:total_disc}
\end{equation} 
\noindent 
%
where $\epsilon_{\rm disc}$ is the observed ellipticity of the disc.

Several parametric laws have been adopted in literature to describe
the surface-brightness distribution of bars.  Ferrers
\citep{laurikainen05},  Freeman \citep{freeman66}, and flat bars   
\citep{prieto97} were considered for the artificial galaxies.

The surface-brightness distribution was assumed to be axially
symmetric with respect to a generalised ellipse
\citep{athanassoula90}.  When the  principal  axes of  the ellipse  are
aligned with the coordinate axes, the radial coordinate is defined as
%
\begin{equation} 
r= \left( |x|^c + 
\left| \frac{y}{(1-\epsilon_{\rm bar})}\right|^c \right)^{1/c}, 
\label{eq:general_ellipse}
\end{equation} 
%
where $\epsilon_{\rm bar}$ is the ellipticity and $c$ controls the
shape of the isophotes. A bar with pure elliptical isophotes has
$c=2$. It is $c>2$ if the isophotes are boxy, and $c<2$ if they are
discy.  The parameters $\epsilon_{\rm bar}$ and $c$ are assumed to be
constant as a function of radius.

The   radial  surface-brightness  profile   of  a   Ferrers  ellipsoid
\citep{ferrers77} is given by
%
\begin{equation} 
I_{\rm bar}^{\rm Ferrers}(r) = \left\{ 
\begin{array}{ll}
  I_{\rm 0,bar} \left(1- \left(\frac{r}{r_{\rm bar}} \right)^2 
    \right)^{n_{\rm bar}+0.5} & \mbox{$r \le r_{\rm bar}$}, \\
  0 & \mbox{$r > r_{\rm bar}$} ,
\end{array} 
\right.
\label{eq:ferrers}
\end{equation} 
%
where $I_{\rm 0,bar}$, $r_{\rm bar}$, and $n_{\rm bar}$ are the
central surface brightness, length, and a shape parameter of the bar,
respectively. The total luminosity is given by
%
\begin{equation} 
L_{\rm bar}^{\rm Ferrers} = 2 \pi I_{\rm 0, bar} r_{\rm bar}^{4}
  \int_{0}^{\infty}r (r_{\rm bar}^{2}-r^{2})^{n_{\rm bar}+0.5} dr, 
\label{eq:total_ferrers_n}
\end{equation} 
%
where $\epsilon_{\rm bar}$ is the ellipticity of the bar. 
The simulated bars were generated by adopting $n_{\rm bar}=2$
following \citet{laurikainen05}.  In this particular case, the total
luminosity of the bar is given by
%
\begin{equation} 
L_{\rm bar}^{\rm Ferrers}= \pi I_{\rm 0,bar} (1-\epsilon_{\rm bar}) 
  r_{\rm bar}^{2}\frac{\Gamma(7/2)}{\Gamma(9/2)}. 
\label{eq:total_ferrers_2}
\end{equation} 

The radial surface-brightness profile of a Freeman bar is
%
\begin{equation} 
I_{\rm bar}^{\rm Freeman} = I_{\rm 0,bar} 
  \sqrt{1-\left(\frac{r}{r_{\rm bar}}\right)^{2}},  
\label{eq:freeman}
\end{equation} 
%
where $I_{\rm 0,bar}$ and $r_{\rm bar}$ are the central surface
brightness and length of the bar, respectively \citep{freeman66}.  The
total luminosity of a Freeman bar is
%
\begin{equation} 
L_{\rm bar}^{\rm Freeman} = \frac{2}{3} \pi(1-\epsilon_{\rm bar}) 
  I_{\rm 0,bar} r_{\rm bar}^{2}, 
\label{eq:total_freeman}
\end{equation} 
%
where $\epsilon_{\rm bar}$ is the ellipticity of the bar.

Finally, the radial surface-brightness profile of a flat bar is
%
\begin{equation} 
I_{\rm bar}^{\rm flat} = I_{\rm 0,bar} 
  \left( \frac{1}{1+e^{\frac{r-r_{\rm bar}}{r_{\rm s}}}} \right), 
\label{eq:flat}
\end{equation} 
%
where  $I_{\rm  0,bar}$ and  $r_{\rm  bar}$  are  the central  surface
brightness and length of the  bar. For radii larger than $r_{\rm bar}$
the surface-brightness profile falls  off with a scale length $r_{\rm
s}$ \citep{prieto97}. The total luminosity is
%
\begin{equation} 
L_{\rm bar}^{\rm flat}= -2\, \pi\, I_{\rm 0,bar}\, (1-\epsilon_{\rm bar})\, 
  r_{\rm s}^{2}\, {\rm Li}_2\left(-e^{r_{\rm bar}/r_{\rm s}} \right), 
\label{eq:total_flat}
\end{equation} 
%
where $\epsilon_{\rm  bar}$ is  the ellipticity of  the bar  and ${\rm
Li}_2$  is  the  dilogarithm  function  (also know  as  the  Jonquiere
function).
 
We generated a set of 8000 images of artificial galaxies with a
S\'ersic bulge and an exponential disc.  Among these galaxies, 2000
have a Ferrers bar, 2000 a Freeman bar, 2000 a flat bar, and 2000 do
not posses a bar.

The  apparent  magnitudes of  the  artificial  galaxies were  randomly
chosen in the range
%
\begin{equation}
10 < m_{r} < 16,
\end{equation}
%
corresponding  to that  of the  sample galaxies.  To  redistribute the
total  galaxy  luminosity  among  the  three  galaxy  components,  the
bulge-to-total  $L_{\rm bulge}/L_{\rm  tot}$  and bar-to-disc  $L_{\rm
bar}/L_{\rm disc}$ luminosity ratio were taken into account. They were
considered to be
%
\begin{equation}
0<L_{\rm bulge}/L_{\rm tot}<0.7, 
\end{equation}
%
and
%
\begin{equation}
0<L_{\rm bar}/L_{\rm disc}<0.3,
\end{equation}
%
following \citet{laurikainen05}.
The adopted ranges for the effective radius of the bulge, scale-length
of the disc, and bar length were selected according to the values
measured for spiral galaxies by \citet{mollenhoff01},
\citet{macarthur03}, \citet{mollenhoff04}, \citet{laurikainen07},
and \citet{mendezabreu08a}. They are
%
\begin{equation}
0.5 < r_{\rm e} < 3\ \mbox{kpc},
\end{equation}
%
\begin{equation}
1 < h < 6 \ \mbox{kpc},
\end{equation}
%
and
%
\begin{equation}
0.5 < r_{\rm bar} < 5\ \mbox{kpc},
\end{equation}
%
respectively. 
The ellipticities  of the structural components were  also selected to
mimic those measured in real galaxies \citep[e.g,][]{marinova07}. They
are
%
\begin{equation}
0.8 < 1-\epsilon_{\rm bulge} < 1,
\end{equation}
%
\begin{equation}
0.5 < 1-\epsilon_{\rm disc} < 1,
\end{equation}
%
and
%
\begin{equation}
0.2 < 1-\epsilon_{\rm bar} <0.7,
\end{equation}
%
with
%
\begin{equation}
\epsilon_{\rm bulge} < \epsilon_{\rm disc} <\epsilon_{\rm bar}.
\end{equation}

Finally,  the position angles  of the  three components  were selected
randomly between $0^\circ$ and  $180^\circ$ to allow each component to
be independently oriented with respect to the others.

In  each pixel  of the  resulting images  noise was  added to  yield a
signal-to-noise ratio  ($S/N$) similar to  that of the  available SDSS
images. It  was given  by the Poisson  noise associated to  the photon
counts due  to both the galaxy  and sky background  and read-out noise
(RON) of the CCD.  The pixel  scale, CCD gain and RON were 0\farcs3946
arcsec   pixel$^{-1}$,  4.72   e$^-$  ADU$^{-1}$,   and   5.52  e$^-$,
respectively. They mimic the instrumental setup of the SDSS images.
In order to  account for seeing effects, the  images of the artificial
galaxies were  convolved with a  Moffat PSF with FWHM=2.77  pixels and
$\beta=3.05$ (see Sect. \ref{sec:sample}).
 
The artificial galaxies do not match the sample galaxies since we did
not account for their redshift distribution. Due to this issue, the
    fractions given in Tab.~\ref{tab:ellipse} and \ref{tab:fourier} do
    not represent at all estimations of the absolute bar fraction lost in real galaxies.  However, they are useful to test the
efficiency of the two proposed methods for detecting bars, in order to
fine tune their free parameters and understand possible biases in the
results.
The artificial galaxies were sized in pixels.  In this way, the
    images are somewhat `dimensionless' and the performances of two
    methods can be assessed by converting the scale lengths from pixel
    to physical units according to the distance of the objects. In
    order to cover the full range of bar lengths and redshifts, the
    shortest bars ($r_{\rm bar} = 0.5$ kpc) were scaled assuming a
    redshift $z=0.04$, while the largest ones ($r_{\rm bar} = 5$ kpc)
    were placed at $z=0.01$.

\subsection{Testing the ellipse fitting method} 
\label{sec:ellipse_test} 

The ellipse fitting method  has two free parameters, $\Delta \epsilon$
and $\rm  \Delta PA$. \citet{laine02} adopted $\Delta  \epsilon = 0.1$
and    $\rm    \Delta     PA    =    20^\circ$,    \citet{marinova07},
\citet{menendezdelmestre07},  and  \citet{barazza08}  adopted  $\Delta
\epsilon = 0.1$ and $\rm \Delta PA=10^\circ$.

We applied  the ellipse fitting  method to the artificial  galaxies by
adopting  $\Delta \epsilon =  0.1, 0.08,  0.05$ and  $\rm \Delta  PA =
10^\circ, 20^\circ, 30^\circ$ to  find the best combination of $\Delta
\epsilon$ and  $\rm \Delta PA$ maximising the  bar identifications and
minimising the bad and/or spurious detections.
The results are given in Table~\ref{tab:ellipse}.

\begin{table} 
\caption{Percentage of galaxies classified as barred and unbarred galaxies erroneously found to be barred by applying the 
ellipse fitting method to the sample of artificial galaxies.}
\label{tab:ellipse} 
\centering 
 \begin{tabular}{c c c c c} 
   \hline\hline 
   \noalign{\smallskip}
   $\Delta$PA & \multicolumn{3}{c}{Barred} &  \multicolumn{1}{c}{Unbarred}  \\
              & Ferrers Bars & Freeman Bars & Flat Bars & \\
   \noalign{\smallskip}
   \hline
   \noalign{\smallskip}
   \multicolumn{5}{c}{$\Delta \epsilon=0.10$}  \\
   \noalign{\smallskip}
   \hline
   \noalign{\smallskip}
   10$^{\circ}$ & 39 & 46 & 22 & 2 \\ 
   20$^{\circ}$ & 53 & 58 & 34 & 3 \\ 
   30$^{\circ}$ & 57 & 63 & 37 & 4 \\
   \noalign{\smallskip}
   \hline
   \noalign{\smallskip}
   \multicolumn{5}{c}{$\Delta \epsilon=0.08$}  \\
   \noalign{\smallskip}
   \hline
   \noalign{\smallskip}
   10$^{\circ}$ & 48 & 55 & 30 & 3 \\
   20$^{\circ}$ & 62 & 67 & 40 & 5 \\          
   30$^{\circ}$ & 66 & 71 & 43 & 6 \\
   \noalign{\smallskip}
   \hline
   \noalign{\smallskip}
   \multicolumn{5}{c}{$\Delta \epsilon=0.05$}  \\
   \noalign{\smallskip}
   \hline
   \noalign{\smallskip}
   10$^{\circ}$ & 65 & 70 & 43 & 9  \\   
   20$^{\circ}$ & 77 & 81 & 54 & 13 \\ 
   30$^{\circ}$ & 79 & 83 & 57 & 15 \\ 
   \noalign{\smallskip}
   \hline 
 \end{tabular} 
\end{table} 

The flat and  Freeman bars are the most difficult  and easiest bars to
be detected, respectively. This  means that the ellipse fitting method
detects  more  efficiently  the   bars  with  sharp  ends  than  those
characterised by a smooth  transition to the disc.  Moreover, adopting
$\rm \Delta PA=20^\circ$ instead of $\rm \Delta PA=10^\circ$ increases
the fraction of bar detections by  $10\%$ for all the bar types, while
the increment between $\rm \Delta PA  = 20^\circ$ and $\rm \Delta PA =
30^\circ$ is only about $4\%$.

Spurious detections correspond to unbarred galaxies which are
erroneously found to be barred. In order to estimate their fraction,
we applied the method to the sample of unbarred artificial galaxies we
built to this aim.  The results for the different values of $\Delta
\epsilon$ and $\rm \Delta PA$ are also given in
Table~\ref{tab:ellipse}.
Bad detections correspond to barred galaxies for which we
  obtained a bad measurement of the bar length.  In order to estimate
  this fraction (which is not reported in Table~\ref{tab:ellipse}),
we compared the bar lengths known for the artificial galaxies with the
$r_{\rm bar}^{\epsilon}$ derived by applying the ellipse fitting
method.
We derived  the median  and standard deviation  of the  relative error
between the known  and measured bar length using  a 3$\sigma$ clipping
iterative procedure. We considered  as bad detections the measurements
with  a relative  error  larger  than 3$\sigma$  with  respect to  the
median.

\begin{figure} 
   \centering 
   \includegraphics[width=9cm]{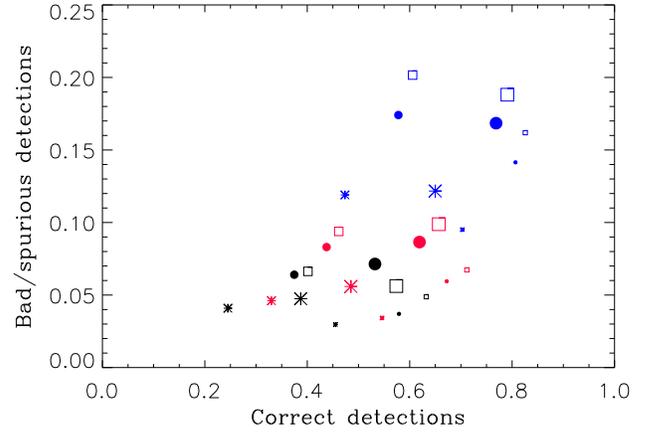} 
      \caption{Fraction of artificial galaxies correctly classified as
              barred vs fraction of bad/spurious bar detections using
              the ellipse fitting method with $\Delta$PA=10$^{\circ}$
              (asterisks), 20$^{\circ}$ (filled circles), and 30$^{\circ}$
              (squares) and $\Delta \epsilon=$0.1 (black symbols),
              0.08 (red symbols), and 0.05 (blue symbols).  The large,
              medium, and small symbols correspond to Ferrers,
              Freeman, and flat bars, respectively.}  
       \label{fig:ellipse} 
   \end{figure}

The fraction of galaxies classified as barred versus the bad/spurious
detections are shown in Fig.~\ref{fig:ellipse} as a function of
$\Delta \epsilon$ and $\rm \Delta PA$.  The optimal configuration is
$\Delta \epsilon=0.08$ and $\rm \Delta PA=20^\circ$ since the fraction
of detections increases by more than $10\%$ with respect to $\Delta
\epsilon=0.1$ and $\rm \Delta PA=10^\circ$, while the fraction of
bad/spurious detections is always lower than $10\%$.   It is worth
  noticing that for $\Delta_\epsilon=0.08$ and $\rm \Delta PA =
  30^\circ$ the fraction of bar detections rises by about $3\%$ with
  respect to $\Delta_\epsilon=0.08$ and $\rm \Delta PA =
  20^\circ$. But the fraction of bad/spurious detections increases
  too. For example, for the Ferrers bars such a fraction is even
  larger than $10\%$.
 
\begin{figure*} 
   \centering 
   \includegraphics[width=15cm]{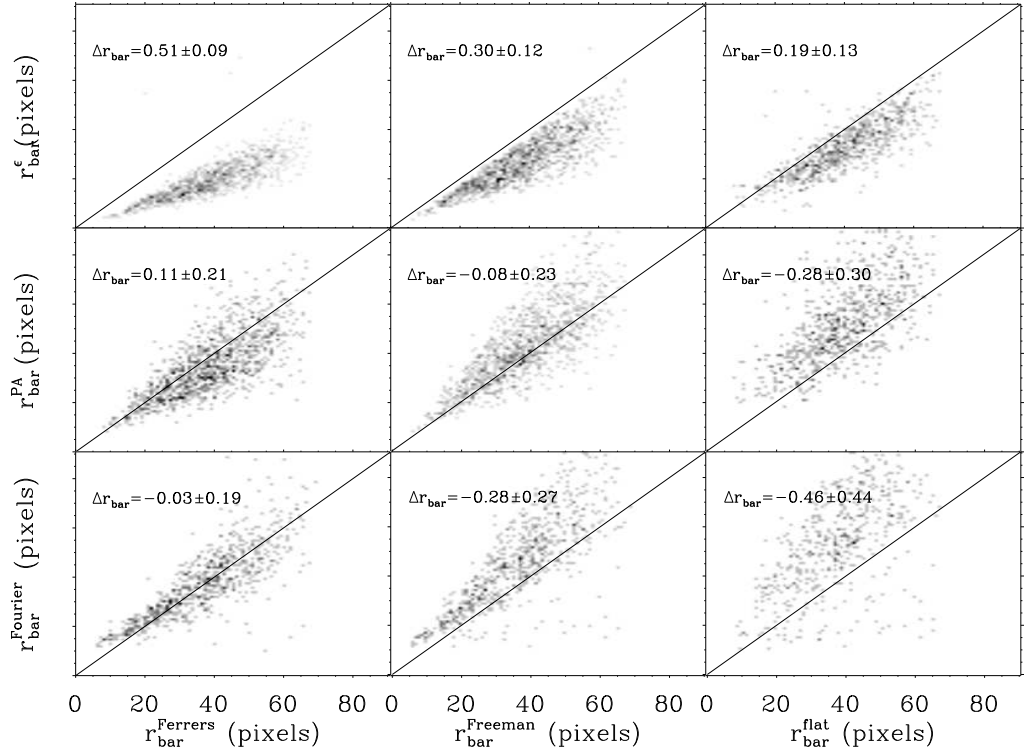} 
      \caption{Bar length measured with the maximum ellipticity (top
        panels), position angle (middle panels), and Fourier analysis
        method (bottom panels) for Ferrers (left panels), Freeman
        (central panels), and flat bars (right panels) in artificial
        galaxies. Mean relative error (defined as the difference
        between input and measured values) and standard deviation for
        the measurements are given in each panel.}
   \label{fig:ellipse_radii} 
   \end{figure*}

The bar  lengths we measured  as $r_{\rm bar}^{\epsilon}$  and $r_{\rm
bar}^{\rm   PA}$   in   the   artificial   galaxies   are   shown   in
Fig.~\ref{fig:ellipse_radii}.  The  bar length is  underestimated when
$r_{\rm  bar}^{\epsilon}$  is   used,  as  found  by  \citet{michel06}
too. This is particularly true for the Ferrers bars where the measured
bar  lengths are  51$\%$ shorter  than  the real  ones.  In  contrast,
Freeman and flat bars were better determined, their measurements being
shorter by 30$\%$ and 19$\%$, respectively.
The bar length  is underestimated when $r_{\rm bar}^{\rm  PA}$ is used
for the Ferrers bars (11$\%$), but it is overestimated for the Freeman
(8$\%$) and flat bars (28$\%$).  These results show the possibility of
define an empirical correction to the bar length, knowing the bar type
in advance.
 
\subsection{Testing the Fourier analysis method} 
\label{sec:fourier_test} 

The deprojection of the galaxy image is a crucial step in applying the
Fourier method, which has  two free parameters, $\Delta (I_{2}/I_{0})$
and $\rm \Delta \phi_{2}$.
The inclination and  major-axis position angle of the  galaxy disc can
be  obtained  by  either  fitting  ellipse  to  the  outermost  galaxy
isophotes \citep[e.g.,][]{aguerri03}  or minimising the  $m=2$ Fourier
mode     in     the     outermost     regions    of     the     galaxy
\citep[e.g.,][]{grosbol85}.  We  applied  these  two  methods  to  the
artificial galaxies  and found that ellipse fitting  gave lower errors
(about $3^\circ$) on both $i$ and PA.
 
\begin{figure} 
   \centering 
   \includegraphics[width=9cm]{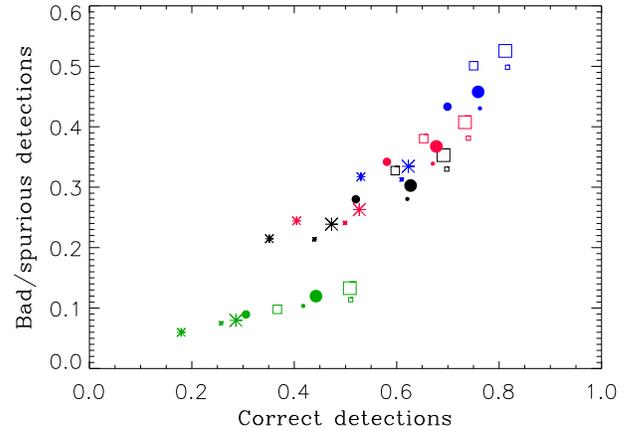}
      \caption{Fraction of artificial galaxies correctly classified as
              barred and fraction of bad/spurious bar detections using
              the Fourier analysis method with $\Delta \phi_2=10^{\circ}$
              (asterisks), 20$^{\circ}$ (filled circles), and 30$^{\circ}$
              (squares) and $\Delta (I_2/I_0)=0.2$ (green symbols),
              0.1 (black symbols), 0.08 (red symbols), and 
              0.05 (blue symbols).  The large, medium, and small symbols 
              correspond to Ferrers, Freeman, and flat bars, respectively.}  
         \label{fig:fourier} 
   \end{figure}

We applied the  Fourier method to the artificial  galaxies by adopting
$\Delta (I_{2}/I_{0}) =0.2, 0.1,  0.08, 0.05$ and $\rm \Delta \phi_{2}
=  10^\circ,  20^\circ, 30^\circ$  to  find  the  best combination  of
$\Delta (I_{2}/I_{0})$  and $\rm \Delta \phi_{2}$,  maximising the bar
identification and minimising the bad and/or spurious detections.
The   results   are    given   in   Table~\ref{tab:fourier}   and   in
Fig.~\ref{fig:fourier}.

\begin{table} 
\caption{Percentage of galaxies classified as barred and unbarred galaxies erroneously found to be barred by applying the 
Fourier analysis method to the sample of artificial galaxies.
}
\label{tab:fourier} 
\centering 
 \begin{tabular}{c c c c c} 
   \hline\hline 
   \noalign{\smallskip} 
   $\Delta \phi_2$ & \multicolumn{3}{c}{Barred} &  \multicolumn{1}{c}{Unbarred}  \\
                   & Ferrers Bars & Freeman Bars & Flat Bars & \\
   \noalign{\smallskip}
   \hline
   \noalign{\smallskip}
   \multicolumn{5}{c}{$\Delta (I_2/I_0)=0.20$} \\ 
   \noalign{\smallskip}
   \hline
   \noalign{\smallskip}
   10$^{\circ}$ & 29 & 26 & 18 & 5 \\
   20$^{\circ}$ & 44 & 42 & 31 & 8 \\
   30$^{\circ}$ & 51 & 51 & 37 & 8 \\
   \noalign{\smallskip}
   \hline
   \noalign{\smallskip}
   \multicolumn{5}{c}{$\Delta (I_2/I_0)=0.10$} \\
   \noalign{\smallskip}
   \hline
   \noalign{\smallskip}
   10$^{\circ}$ & 47 & 44 & 35 & 19 \\
   20$^{\circ}$ & 63 & 62 & 52 & 26 \\
   30$^{\circ}$ & 69 & 70 & 60 & 31 \\
   \noalign{\smallskip}
   \hline
   \noalign{\smallskip}
   \multicolumn{5}{c}{$\Delta (I_2/I_0)=0.08$} \\
   \noalign{\smallskip}
   \hline
   \noalign{\smallskip}
   10$^{\circ}$ & 53 & 50 & 40 & 22 \\
   20$^{\circ}$ & 67 & 67 & 58 & 32 \\
   30$^{\circ}$ & 73 & 74 & 65 & 36 \\
   \noalign{\smallskip}
   \hline
   \noalign{\smallskip}
   \multicolumn{5}{c}{$\Delta (I_2/I_0)=0.05$} \\ 
   \noalign{\smallskip}
   \hline
   \noalign{\smallskip}
   10$^{\circ}$ & 62 & 61 & 53 & 29 \\ 	
   20$^{\circ}$ & 75 & 76 & 70 & 41 \\ 
   30$^{\circ}$ & 81 & 82 & 75 & 48 \\ 
   \noalign{\smallskip}
   \hline 
 \end{tabular} 
\end{table} 

In general,  the Fourier  method is less  efficient in  detecting bars
than the ellipse fitting method. We found that $\Delta (I_{2}/I_{0}) =
0.2$ has to  be adopted to have a  fraction of bad/spurious detections
lower  than  $10\%$. We  also  adopted  $\Delta  \phi_{2} =  20^\circ$
because it  increases detections  by more than  $10\%$ and  gives less
bad/spurious detections with respect to $\Delta \phi_{2} = 30^\circ$.
The  bar lengths  we measured  as $r_{\rm  bar}^{\rm Fourier}$  in the
artificial  galaxies are  shown  in Fig.~\ref{fig:ellipse_radii}.  The
method recovers the bar length  with the best accuracy for the Ferrers
bars (3$\%$  error), while  the bar length  is over estimated  for the
Freeman (28$\%$) and flat bars (46$\%$).

\section{Bar fraction} 
\label{sec:fraction} 

Both the ellipse fitting and Fourier method were applied to our sample
of 2106 disc galaxies.
We found  that the fraction  of galaxies classified as  barred depends
strongly on the technique adopted  for the analysis: it is $45\%$ with
the ellipse fitting method and $26\%$ with the Fourier method.
Although the Fourier method was demonstrated to be less efficient than
ellipse fitting in detecting bars, this difference is larger than that
expected from the analysis of the artificial galaxies.
To  investigate this  issue, we  took into  account  the morphological
classification of  the galaxies found  to be barred. According  to the
ellipse  fit method  $29\%$,  $55\%$, and  $54\%$  of the  lenticular,
early-type  and late-type spiral  galaxies, respectively,  are barred.
They are  $29\%$, $33\%$, and  $17\%$, respectively, with  the Fourier
method.   Therefore, both methods  obtained a  similar of  fraction of
barred lenticular galaxies, while the Fourier method is less efficient
in  detecting bars in  spirals, and  particularly in  late-type spiral
galaxies.

\begin{figure} 
   \centering 
   \includegraphics[width=9cm]{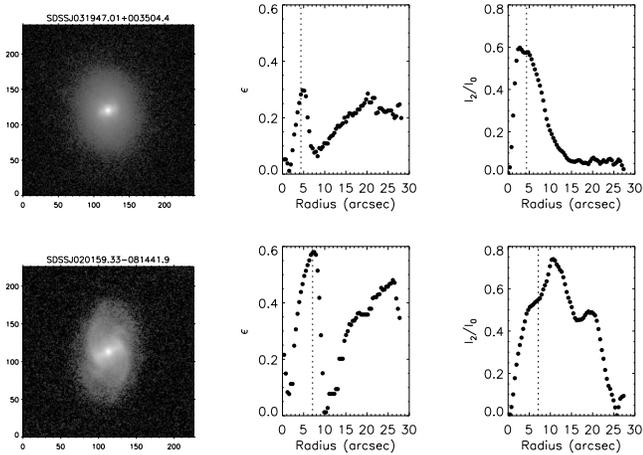} 
      \caption{The $r-$band images (left panels), ellipticity 
         (central panels), and $m=2$ Fourier amplitude radial profiles
         (right panels) of the early-type spiral
         SDSSJ031947.01$+$003504.4 (top panels) and  the late-type
         spiral SDSSJ020159.33$-$081441.9 are analysed by both methods 
         (bottom panels). For each galaxy the vertical
         dotted line corresponds to the value of $r_{\rm bar}^{\epsilon}$.}
      \label{fig:examples} 
   \end{figure}

An example is shown in Fig.~\ref{fig:examples}. 
The  early-type  spiral  SDSSJ031947.01$+$003504.4 and  the  late-type
spiral SDSSJ020159.33$-$081441.9 are analysed by both methods.
The bar of the early-type spiral  was detected by both methods and the
measured bar lengths are in agreement. In fact, the radial profiles of
$\epsilon$  and $I_{2}/I_{0}$  show a  local maximum  at  about $5''$,
where the PA and $\phi_2$ are constant.
On the contrary, the bar of  the late-type spiral was detected only by
the  ellipse  fit  method.   The  radial profiles  of  $\epsilon$  and
$I_{2}/I_{0}$ show a local  maximum at different radii (about $10''$).
The $I_{2}/I_{0}$ maximum  is located in the spiral  arm region, where
$\phi_2$ is not  constant.  Therefore, the bar of  this galaxy was not
detected by the Fourier method.
We conclude  that bars  with sharp ends  are detected by  both ellipse
fitting and Fourier methods. But,  the bars of galaxies with lenses or
strong spiral arms  are more easily detected with  the ellipse fitting
method.   This kind  of bars  is usually  found in  late-type spirals.
These large differences  in the bar fractions between  the two methods
could  bias our  conclusions.   For  this reason,  we  will study  the
photometrical  parameters of  the bars  by adopting  only  the ellipse
fitting method.

As an additional check, we visually classified all galaxies in barred
and unbarred. The visual classification was done by two of us
  (JALA and JMA).  Both the classifications were in close agreement
  and only their mean is reported. No attempt to classify the galaxies
  according to their Hubble type was done. The difference between the
  bar fractions found with the visual and the automatic classification
  ($\sim 7\%$) is our best estimate of the fraction of undetected bars
  in the galaxy sample. We obtained that the global fraction of
  barred galaxies in our sample was 38$\%$. Taking into account the
  different morphological types we found that 22$\%$, 52$\%$ and
  48$\%$ of the lenticular, early-type and late-type spiral galaxies,
  respectively, were barred.

Our bar fraction ($45\%$) of disc galaxies in the local universe is in
good  agreement  with recent  results  obtained  in  optical bands  by
\citet[][$44\%$ in  the $B$ band]{marinova07},  \citet[][$47\%$ in the
$I$ band]{reese07}, and \citet[][$50\%$ in the $r$ band]{barazza08}.
In  addition, our  finding that  early- and  late-type spirals  host a
larger  fraction  of  bars  than  lenticular  galaxies,  was  also  in
agreement with \citet{barazza08}  since they found that disc-dominated
galaxies with low bulge-to-disc  luminosity ratio display a higher bar
fraction   than   galaxies    with   significant   bulges.    However,
\citet{marinova07} did not find any  difference in the bar fraction in
the NIR as  a function of the Hubble type, we  argue that their result
is  biased by  their smaller  coverage  of the  Hubble sequence  since
neither lenticulars  nor Sd/Sm galaxies  were taken into  account. The
same   consideration  can  be   applied  to   the  results   found  by
\citet{knapen00} and \citet{eskridge00}.

\section{Bar properties} 
\label{sec:bar_properties}

\subsection{Bar length} 
\label{sec:length} 

\begin{figure*} 
   \centering 
   \includegraphics[width=15cm]{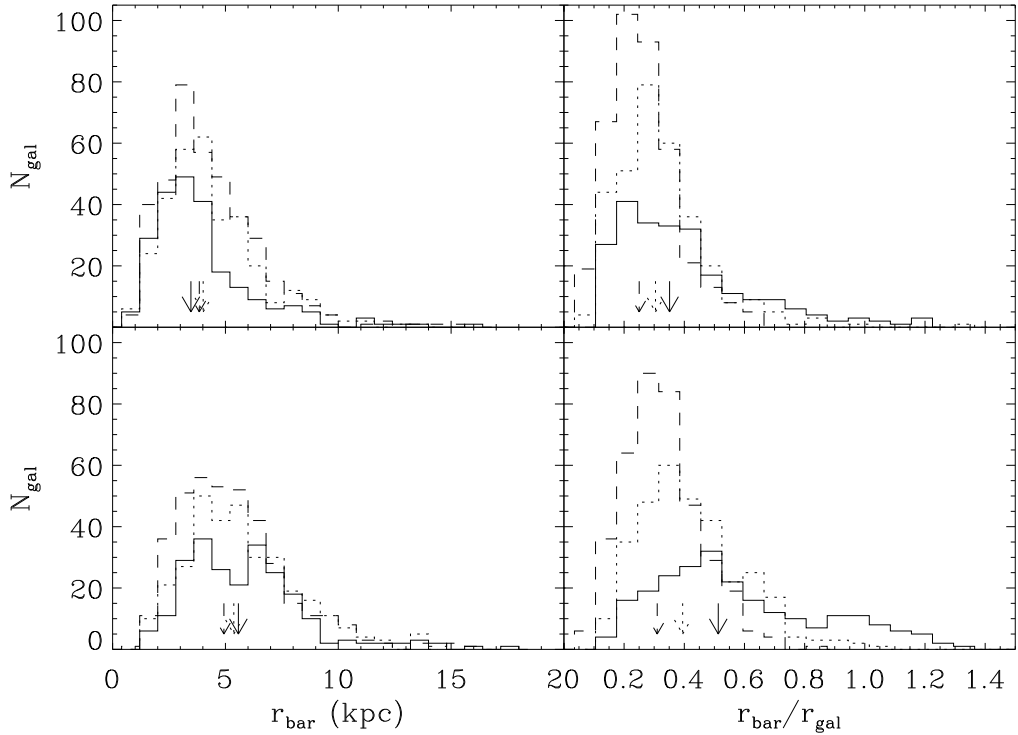} 
      \caption{Distribution of the bar length (left  panels) and 
               normalised bar length (right panels) in lenticulars
               (full line), early-type spirals (dotted line), and
               late-type spirals (dashed line). The bar length was
               measured with both the maximum ellipticity (top
               panels) and PA method (bottom panels). Arrows mark
               the median values of the distributions.}
              \label{fig:length} 
   \end{figure*}

The distributions of  the bar lengths and normalised  bar lengths were
derived  for the  sample  galaxies after  deprojection  on the  galaxy
plane.  They   are  shown  in   Fig.~\ref{fig:length}.   Both  $r_{\rm
bar}^{\epsilon}$ and  $r_{\rm bar}^{\rm PA}$ were  considered, and the
galaxy size was  defined as $r_{\rm gal} = 2  \times r_{\rm p}$, where
$r_{\rm p}$ is the Petrosian radius from SDSS.
The median values derived for the different morphological bins are
given in Tab.~\ref{tab:length}.  The values of $r_{\rm
    bar}^{\epsilon}$ are systematically smaller than those of $r_{\rm
    bar}^{\rm PA}$, as expected from the measurements of the
  artificial galaxies.
The comparison of our results, with previous works where the $r_{\rm bar}^{\epsilon}$ values are reported, gives us a good agreement. For example, Erwin (2005) found a median bar length of 3.3 kpc, Marinova \& Jogee (2007) calculate a mean value of 4 kpc and Menendez-Delmestre et al. (2007) obtain a median value of 3.5 kpc. These results hold even considering only the bars with a length
larger than 2 kpc. This limit corresponds to minimum bar length we are
able to resolve all throughout our range of distances.

\begin{table} 
\caption{Median values of the bar radius for the different galaxy types.} 
\label{tab:length} 
\centering 
\begin{tabular}{l c c c}     
\hline\hline 
\noalign{\smallskip}                      
  & S0 & Early-type spirals & Late-type spirals \\
\noalign{\smallskip}                       
\hline 
\noalign{\smallskip}                      
$r_{\rm bar}^{\epsilon}$ (kpc)     & 3.5 & 4.0 & 3.8 \\  
$r_{\rm bar}^{\rm PA}$ (kpc)             & 5.6 & 5.4 & 4.9 \\ 
$r_{\rm bar}^{\epsilon}/r_{\rm gal}$ & 0.35 & 0.30 & 0.25 \\ 
$r_{\rm bar}^{\rm PA}/r_{\rm gal}$         & 0.51 & 0.39 & 0.31 \\ 
\noalign{\smallskip}                      
\hline 
\end{tabular} 
\end{table} 

\begin{figure} 
   \centering 
   \includegraphics[width=9cm]{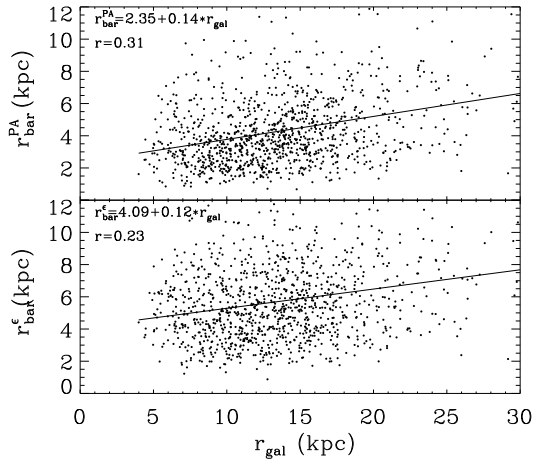} 
      \caption{Galaxy radius $r_{\rm gal}$ versus bar length measured
        with the PA method (top panel) and maximum ellipticity method
        (bottom panel). The solid line represents the linear
        regression through all the data points. The Pearson
        correlation coefficient ($r$) and the result of the linear fit
        are given.}
         \label{fig:length_size} 
   \end{figure}

Since the bar length is strongly dependent on the method used to
    derive it, we can not conclude much about the correlation between
    the bar length and the morphological type. According to $r_{\rm
      bar}^{\epsilon}$, the lenticulars host the shortest bars, while
    according to $r_{\rm bar}^{\rm PA}$ their bars are the longest
    ones. As far as the median bar length of the spirals concerns, the
    late-type spirals host shorter bars with respect to the early-type
    ones (Tab.~\ref{tab:length}).
Nevertheless, a correlation between the bar length and galaxy size was
found.  Thus, larger bars are located in bigger galaxies
(Fig.~\ref{fig:length_size}).  The correlation is independent of the
adopted method to measure the bar length. It holds for the different
morphological bins too, being stronger for late-type spirals
($r=0.52$) and weaker for S0 galaxies ($r=0.38$). This relation could
indicate a link between the formation and evolution processes between
of bars and galaxy discs.  A similar correlation was found by
\citet{marinova07}, although a quantitative comparison with them is
not possible due to the different band-passes and different definition
of the galaxy radius they adopted.

\subsection{Bar strength} 
\label{sec:strength} 
 
The bar strength  represents the contribution of the  bar to the total
galaxy potential.   Several methods have been developed  to measure it
\citep[see][and references therein]{laurikainen07}. Nowadays, the most
commonly used parameter measuring the bar strength is $Q_{\rm g}$
defined by \citet{buta01}. It can be accurately estimated by analysing
near-infrared images \citep[][]{buta03, block04, laurikainen07}, which
are not available for our sample galaxies drawn from the SDSS.
However, \citet{abraham00} defined another bar strength parameter given
by
%
\begin{equation} 
f_{\rm bar} = \frac{2}{\pi} \left(\arctan (1-\epsilon_{\rm bar})^{-1/2}
  - \arctan (1-\epsilon_{\rm bar})^{+1/2} \right), 
\end{equation}
%
where  $\epsilon_{\rm  bar}$  is   the  bar  ellipticity  measured  at
$r^{\epsilon}_{\rm   bar}$.    It    correlates   with   $Q_{\rm   g}$
\citep[e.g.,][]{laurikainen07}   and  was   adopted  for   our  sample
galaxies.

\begin{figure} 
   \centering  
   \includegraphics[width=9cm]{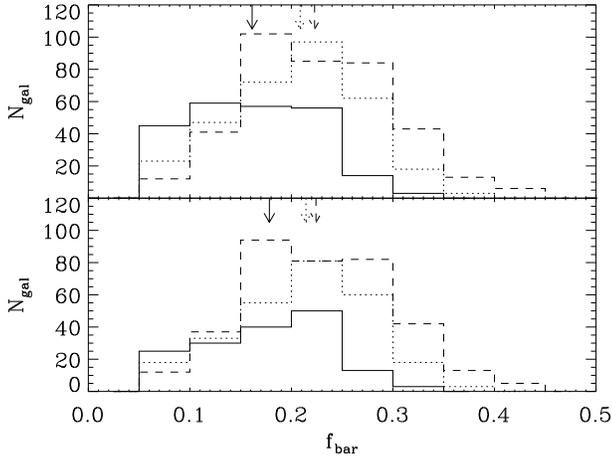} 
      \caption{Distribution of the bar strength in lenticulars (full
        line), early-type spirals (dotted line), and late-type spirals
        (dashed line)  for the whole sample (top panel) and for
          the galaxies with $r_{\rm bar} > r_{\rm bd}$ (bottom
          panel). Arrows mark the median values of the
        distributions.}
   \label{fig:strength} 
   \end{figure}

We did not adopted any minimum value for the bar ellipticity.  We
    found $\epsilon_{\rm bar, min}=0.16$, which is close the minimum
    ellipticity adopted in other studies \citep[e.g., $\epsilon_{\rm
        bar, min}=0.2$,][]{marinova07}.
The distributions  of the bar  strengths we derived for  the different
morphological types are shown in Fig.~\ref{fig:strength}.
The median values for the bar strengths of the lenticular, early-type,
and late-type spiral galaxies  are 0.16, 0.19, and 0.20, respectively.
Indeed, we  found a significant difference between  the lenticular and
spiral galaxies. They are characterised by different distributions, as
confirmed at a high confidence level ($>95\%$) by a Kolmogorov-Smirnov
(KS) test.  Using  four different methods to derive  the bar strength,
\citet{laurikainen07} also  found that S0  galaxies host significantly
weaker bars than the rest of  disc galaxies, this result was hold also
by  \citep[][]{das03}   and  \citet{barazza08}  using   only  the  bar
ellipticity.    In  contrast,   \citet{marinova07}   found  that   the
ellipticity  of  the bar  is  practically  independent  of the  Hubble
type.  But they  not consider  S0  galaxies which  are those  actually
making the difference.

However, it could be possible that the presence of a large bulge
    could affect the measurement of the bar ellipticity, and therefore
    the calculation of the bar strength.  In order to address this
    issue, we performed a further test.  We fitted an exponential law
    to the outer parts of the surface-brightness profiles of our
    barred galaxies.  Then, we computed the radius $r_{\rm bd}$ at
    which the galaxy surface brightness profile exceeds the fitted
    exponential. This radius represents an estimate of the extension
    of the region where the bulge contribution dominates the light of
    the galaxy.  At this point, we selected a subsample of barred
    galaxies with $r_{\rm bar} > r_{\rm bd}$. We recalculated the mean
    strength of these bars by splitting the sample in lenticulars,
    early- and late-type spirals.  As expected, in the new subsample
    of galaxies we lost the weakest bars, especially in the lenticular
    galaxies.  Nevertheless, the final result is the same: the
    lenticulars have weaker bars than the early- or the late-type
    spirals.

Some numerical simulations of bar formation and evolution propose that
bars can  be formed and destroyed  fastly due to the  accretion of gas
towards  the central regions  of the  galaxies \citep[][]{pfenniger90,
bournaud02,  bournaud05}.  In  this  framework, due  to the  fast
destruction  and re-formation  of bars,  we would  expect  a bimodal
distribution of  the bar strength at least  for gas-rich galaxies
like sthe late-type barred ones. The absent of this bimodality in the
bar strength  for all  galaxy types showed  in Fig.~\ref{fig:strength}
could be against those bar formation and evolution scenarios.
 
\section{Bars and galaxy properties} 
\label{sec:galaxy_properties}

\subsection{Galaxy local environment} 
\label{sec:environment}
 
Due  to   our  selection  criteria   we  excluded  all   the  strongly
disturbed/interacting galaxies.  Nevertheless, we calculated  for each
sample  galaxy  the  local  density  following  the  prescriptions  of
\citet{balogh04a,  balogh04b}  in order  to  investigate the  relation
between the bar  properties and local environment of  the host galaxy.
The number density  of local galaxies was computed  using the distance
$d_{5}$ of the galaxy to  its fifth nearest neighbour galaxy.  Thus, a
projected galaxy density could be defined as
%
\begin{equation}
\Sigma_5 = \frac{5}{\pi d_5^2} .
\end{equation}

This was computed  with those galaxies located in  a velocity range of
$\pm   1000$   km  s$^{-1}$   from   the   target   galaxy  to   avoid
background/foreground  contamination.  For  sample galaxies  without a
measured redshift we imposed a  luminosity constraint of $\pm3$ mag to
derive the galaxy density as done by \citet{balogh04a, balogh04b}.

\begin{figure} 
   \centering 
   \includegraphics[width=9cm]{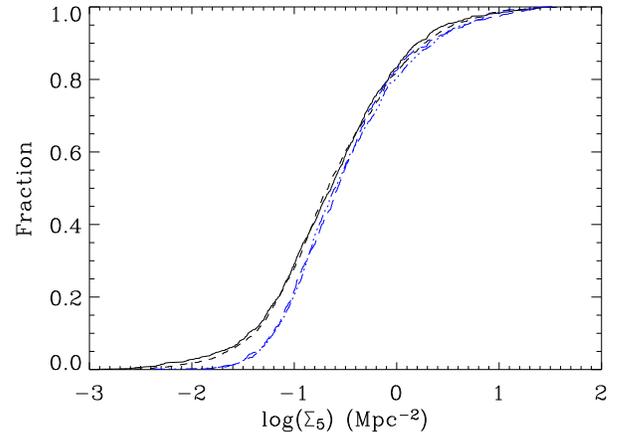} 
      \caption{Cumulative distribution of barred
      (solid black line) and unbarred (dashed black line) galaxies as
      function of the local galaxy density. Cumulative
      distribution of barred (long-dashed blue line) and unbarred
      (dashed-dotted blue line) galaxies after excluding those at less
      than 7 Mpc from the nearest edge of SDSS is also overplotted (see text for more details).}
   \label{fig:local_density} 
   \end{figure}

Fig.~\ref{fig:local_density} shows the fraction of barred and unbarred
galaxies in our sample as function of the local galaxy density. In the
range of  galaxy density covered by  our sample, there  is no relation
between the presence of a bar  and the environment of the host galaxy.
The same is true even  if the galaxies of different morphological type
are  considered  independently.  In  addition,  we did  not  find  any
correlation between  the bar length  or strength and the  local galaxy
density.

However,  a series  of caveats  must be  taken into  account  when the
distance neighbour method is applied.

\begin{figure} 
   \centering 
   \includegraphics[width=9cm]{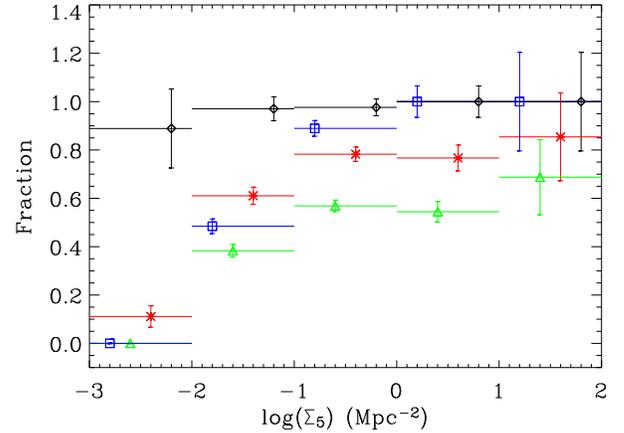} 

      \caption{Fraction of galaxies remaining in the sample after
        correcting for survey edge effects by removing all the
        galaxies with an edge of the survey closer than the measured
        fifth neighbour distance (blue squares), 2 Mpc (black
        diamonds), 5 Mpc (green triangles), and 7 Mpc (red
        asterisks). Poissonian errors are given. Horizontal lines show
        the width of the local density bins. }
   \label{fig:density_border} 
   \end{figure}

The limited area of sky covered by the survey implies that usually the
estimated density is  lower than the true one.  In  fact, due to edges
and holes in  the survey, the value of  $d_{5}$ could be overestimated
and the  derived density underestimated.  Two methods were  applied to
our sample in  order to test the robustness of  our result against the
edge effects.  The first consists  in removing all the  galaxies whose
distance  to the  survey  edges  is smaller  than  the measured  fifth
neighbour distance \citep{miller03, balogh04b}. In this way, we ensure
that the remaining  galaxies have an unbiased estimation  of the local
density.  The fraction  of galaxies  we measured  in each  density bin
after     applying     such    a     correction     is    shown     in
Fig.~\ref{fig:density_border}. It is worth  noticing that no galaxy is
available  in  the  bin   of  lowest  local  density  ($\Sigma_5<0.01$
Mpc$^{-2}$).  This  means that such a  low density can  not be derived
for any  galaxy of the sample.  On the contrary, the  local density is
correctly  derived for  all the  sample galaxies  with $\Sigma_5  > 1$
Mpc$^{-2}$.

Unfortunately, this  method biases the  distribution toward over-dense
environments. This bias  can be reduced by excluding  all the galaxies
within  a given  distance  to the  nearest  edge \citep{cooper05}.  We
excluded the  galaxies at less than 2,  5, and 7 Mpc  from the nearest
edge  of SDSS (Fig.~\ref{fig:density_border}).  If the  lowest density
bin  is excluded,  the  method introduces  only  a weak  contamination
toward high density  environments. An optimal given distance  of 7 Mpc
was found by  calculating the maximum distance to  the fifth neighbour
in  the  bin  of lowest  density.   In  this  way, the  local  density
measurements are reliable and there is no bias toward the high-density
environments,  i.e.  all the  bins  have  almost  the same  number  of
galaxies.
Fig.~\ref{fig:local_density} shows the fraction of barred and unbarred
galaxies in our  sample as function of the  local galaxy density after
excluding  galaxies  at less  than  7 Mpc  from  the  nearest edge  of
SDSS.  They  are  about $40\%$  of  the  total.  We  do not  find  any
difference between the environment of barred and unbarred galaxies.

Also the selected redshift range could lead to underestimate the local
density. We  circumvent this problem by defining  a new volume-limited
sample in a wider redshift range ($0 < z < 0.06$) taking into account
the adopted velocity range.

Finally, the local density distribution  could be biased by SDSS fiber
collision which prevents to  measured galaxies closer than $55''$ with
respect to each  other. In SDSS-DR5 the net  effect of fiber collision
is a loss  of $6\%$ of the galaxies in  the photometric catalogue that
would     otherwise    be     in    the     spectroscopic    catalogue
\citep{cowanivezic08}.  In our  case, this  value represents  an upper
limit since all the galaxies  in photometric catalogue were taken into
account in calculating the local density.

According  to numerical simulations,  galaxy mergers  and interactions
are   mechanisms   which   should   drive  the   formation   of   bars
\citep[][]{gerin90,  miwa98,   mastropietro05}.  Therefore,  we  could
expect  that fraction  of  barred galaxies  increases  with the  local
density.   But, the observational  proofs about  the influence  of the
environment  on bar  formation  and evolution  are  few. For  example,
\citet{thompson81} suggested  a link  between bar formation  and local
galaxy environment  by observing that the fraction  of barred galaxies
increases towards the core of Coma  cluster.  But, this is not case in
the  wide range  of densities  we explored  ($0.01 <  \Sigma_5  < 100$
Mpc$^{-2}$, Fig.~\ref{fig:local_density}). For  the lowest density bin
our fraction of  barred galaxies is even smaller  than 60$\%$ found by
\citet{verley07}  by  analysing  the  optical images  of  45  isolated
galaxies. Recently, \citep[][]{marinova08} shows that the cluster environment does not strongly affect to the bar fraction.
We  argue  that  for  non-interacting  and  undisturbed  galaxies  the
environment do not play a major role in the formation and evolution of
their bar.

Fig.~\ref{fig:local_density} shows that $80\%$ of the sample
    galaxies are located in very low-density environments ($\Sigma_5 <
    1$ Mpc$^{-2}$). The local density of the remaining $20\%$ (corresponding to more than
400 galaxies) covers mostly typical values measured for loose
($\Sigma_5 > 1$ Mpc$^{-2}$) and compact galaxy groups ($\Sigma_5 \sim
10$ Mpc$^{-2}$).  Nevertheless, the fraction of barred galaxies does
    not depend on the local density also for these
    galaxies. Therefore, we conclude that the environment does not
    play an important role in the formation of bars, at least over the
    observed range of local densities. Moreover, it does not account
    for the variation of the central light concentration and galaxy
    colours discussed in Sects. \ref{sec:concentration} and
    \ref{sec:colors}, respectively. Similarly, low  density environments, as those reported here,
do not also account for variations in other galaxy properties, such as
the blue  galaxy fraction \citep[e.g.,][]{aguerri07}. However, we can not infer that
    close interactions do not affect bar formation and evolution,
    because we selected only non-strongly disturbed/interacting
    galaxies.

 This result is in agreement with the numerical simulations by
  \citet[][]{heller07, romanodiaz08}, who argue that there is no
  difference between the bar fraction for field and cluster
  galaxies. They claim that the bar evolution is mainly driven by the
  dark matter subhalos, which surround all the bright galaxies and do
  not depend of their environment. These subhalos could host faint
  galaxies, which are not visible in our images.

\subsection{Central light concentration} 
\label{sec:concentration}

\begin{figure} 
   \centering     
   \includegraphics[width=9cm]{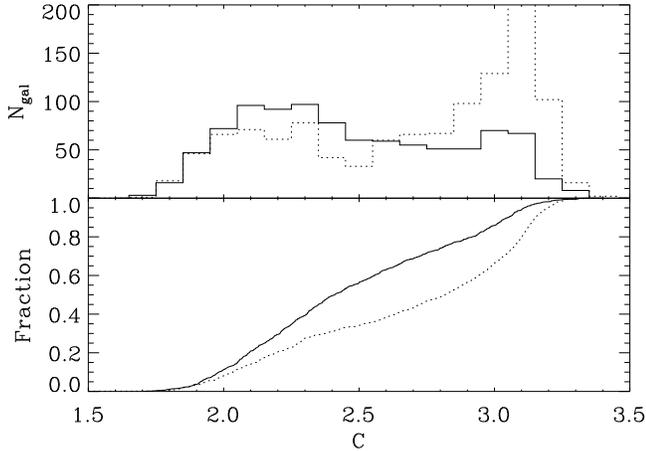}
   \caption{Number  of  barred  (full  line) and  unbarred  disc galaxies
      (dotted  line) as  function  of the  light concentration  (upper
      panel).  Cumulative  distribution  of  barred  (full  line)  and
      unbarred  disc galaxies (dashed-line)  as   function  of  the  light  galaxy
      concentration (lower panel). }
   \label{fig:concentration} 
   \end{figure}

Fig.~\ref{fig:concentration}  shows   the  distribution  of   the  $C$
parameter  for  the barred  and  unbarred  sample  galaxies and  their
cumulative  distribution  functions.    Both  types  of  galaxies  are
characterised  by  different  distributions  as confirmed  at  a  high
confidence level ($>99\%$ ) by a  KS test. It is worth noting that the
differences between both distributions  is due to galaxies with higher
central light  concentration. This result  holds even if we  take into
account  the contamination  of  ellipticals into  our  sample of  disc
galaxies (Sect. \ref{sec:fraction}).

\citet{barazza08} found that the fraction of barred galaxies is higher
for the galaxies with a smaller value of the S\'ersic parameter (i.e.,
the     less-concentrated      galaxies).     We     confirm     their
findings. Fig.~\ref{fig:concentration} shows that the number of barred
and  unbarred galaxies  is clearly  different for  galaxies  with high
values of $C$, being the  fraction of barred galaxies smaller than the
unbarred ones.

Since  light concentration  is  correlated with  the central  velocity
dispersion, the  previous result implies  that, in some  way, galaxies
with higher central mass  concentrations tend to inhibit the formation
and/or evolution of bars. This is in agreement with the results of the
numerical    experiments    by    \citet[][]{pfenniger90,    norman96,
athanassoula02, debattista06}, who showed that the presence of a large
bulge weakens the bar.

\subsection{Galaxy colours} 
\label{sec:colors}

\begin{figure} 
   \centering 
   \includegraphics[width=9cm]{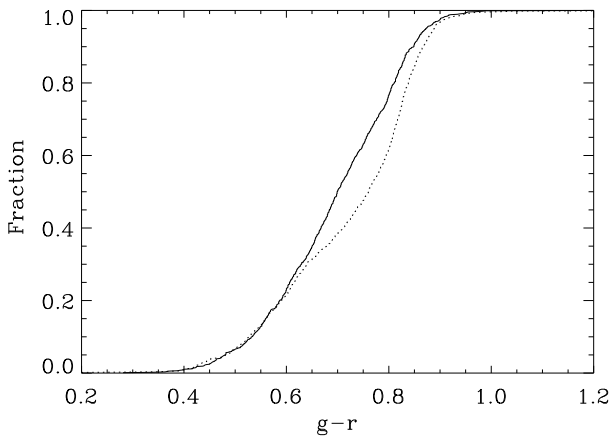} 
      \caption{Cumulative distribution of barred (solid line) 
      and unbarred (dashed line) as function of the $g-r$ galaxy 
      colour.} 
   \label{fig:colors} 
   \end{figure}

Fig.~\ref{fig:colors} shows  the cumulative distribution  functions of
the $g-r$ colour for the sample galaxies. Barred and unbarred galaxies
are characterised  by different distributions  as confirmed at  a high
confidence level  ($>99\%$) by  a KS test.  Thus, barred  galaxies are
bluer than unbarred  ones. We can explained this effect  as due to the
larger fraction of barred  galaxies observed in the late-type systems,
which  are  systematically bluer  than  the  early-type ones.  Similar
colour  difference  between  barred  and unbarred  galaxies  was  also
reported by \citet{barazza08}.

\section{Conclusions} 
\label{sec:conclusions}

We have  studied the fraction  and properties of  bars in a  sample of
$2106$  disc  galaxies  extracted   from  the  SDSS-DR5.   This  is  a
volume-limited sample  of undisturbed and  non-interacting galaxies in
the  redshift range  $0.01 <  z <  0.04$, with  an  absolute magnitude
$M_{r} < -20$, and an inclination $i<60^{\circ}$.

The  sample galaxies  have been  classified lenticulars,  early- (i.e,
S0/a -- Sb)  or late-type spirals (i.e, Sbc --  Sm) according to their
light   concentration.   The  light   concentration  was   defined  as
$C=r_{90}/r_{50}$, where $r_{50}$ and $r_{90}$ are the radii enclosing
$50\%$  and $90\%$  of the  total galaxy  light,  respectively.  These
radii  are available  in  the SDSS  database  for all  objects of  our
sample.    The  correlation  between   the  light   concentration  and
morphological type was derived from a subsample of 612 galaxies listed
in RC3, whose morphological classification was already known.
We  found  that the  fraction  of  lenticulars,  early- and  late-type
spirals among the selected disc galaxies is $39\%$, $28\%$, and $33\%$
respectively.

We  derived the  fraction of  barred  galaxies by  analysing the  SDSS
$r-$band images with the ellipse fitting and Fourier analysis methods.
They consist in  looking for a local maximum in  the radial profile of
ellipticity   \citep[associated   to   a  constant   position   angle;
e.g.,][]{wozniak95}    and    $m=2$    relative   Fourier    component
\citep[associated  to a  constant  phase angle;  e.g.,][]{aguerri00a},
respectively.

The bar  fraction depends  strongly on the  technique adopted  for the
analysis.
By  extensive  testing on  a  large  set  of artificial  galaxies,  we
concluded that  the both methods  are efficient in detecting  the bars
with  sharp  ends, such  as  the Ferrers  and  Freeman  bars.  On  the
contrary,  the  flat  bars,   which  are  characterised  by  a  smooth
transition  to the  disc,  are  more easily  detected  by the  ellipse
fitting method.
The  ellipse fitting  method is  more efficient  in detecting  bars in
galaxies with lenses and spiral arms, where the $m=2$ relative Fourier
component  shows  multiple maxima  (but  it  is  not associated  to  a
constant phase angle). This is the case for late-type spiral galaxies.
For this  reason, we  decided to rely  only onto the  results obtained
with the ellipse fitting method.  We found that $45\%$ of the selected
disc  galaxies host  a  bar  in agreement  with  previous findings  in
optical wavebands \citep{marinova07, reese07}. The fraction of bars in
the three  different morphological bins is $29\%$,  $55\%$, and $54\%$
for  lenticulars,  early-  and  late-type spirals,  respectively.   By
classifying visually  the galaxies in barred and  unbarred we obtained
similar bar fractions as those reported by the ellipse fitting method.

The  bar   length  was  obtained  by  measuring   the  radius  $r_{\rm
bar}^{\epsilon}$ at  which the maximum ellipticity was  reached and as
the radius $r_{\rm bar}^{\rm PA}$ at which the PA changes by $5^\circ$
with  respect to the  value corresponding  to the  maximum ellipticity
\citep[e.g.,][]{wozniak95}.
According to the  analysis of the artificial galaxies,  the bar length
is underestimated  when $r_{\rm bar}^{\epsilon}$ is used,  as found by
\citet{michel06} too.  This is  particularly true for the Ferrers bars
where the measured bar lengths  are 51$\%$ shorter than the real ones.
In  contrast, Freeman  and  flat bars  were  better determined,  their
measurements being shorter by 30$\%$ and 19$\%$, respectively.
The bar length  is underestimated when $r_{\rm bar}^{\rm  PA}$ is used
for the Ferrers bars (11$\%$), but it is overestimated for the Freeman
(8$\%$) and flat bars (28$\%$).  These results show the possibility of
define an empirical correction to the bar length, knowing the bar type
in advance.
We obtained  that the bar  lengths (when normalised by  the galaxy
size) are larger in lenticular galaxies than those presented in early-
and late-type ones. This finding is independent of the method used for
measuring the bar length, and statistically significant according with
the KS  test. We also found  a correlation between the  bar length and
galaxy size.  This correlation is  also independent of the method used
for measuring the bar length. It holds for the different morphological
bins, being  stronger for late-type spirals ($r=0.52$)  and weaker for
lenticular galaxies ($r=0.38$). The  larger bars are located in larger
galaxies, indicating an  interplay between the bar and  disc in galaxy
evolution.

The   bar  strength   $f_{\rm  bar}$   was  estimated   following  the
parametrisation by  \citet{abraham00} which requires  the measurement
of the bar ellipticity.
The median values for the bar strengths of the lenticular, early-type,
and late-type spiral galaxies are 0.16, 0.19, and 0.20, respectively.
The bars of the lenticular galaxies were found to be weaker than those
in spirals, as  found by \citet{laurikainen07} too. The  fact that the
bar  strength distribution are  unimodal for  all galaxy  types argues
against  evolutionary  models  in  which  bars  would  be  formed  and
destroyed in short timescales.

No  difference between  the  local galaxy  density  was found  between
barred  and  unbarred  galaxies   in  our  sample.   Thus,  the  local
environment  does  not seem  to  influence  bar formation.   Moreover,
neither the  length nor strength of  the bars are  correlated with the
local galaxy environment. The  previous results are even true for
the subsample of  our galaxies located in the  more dense environments
(log($\Sigma_5)>0$ Gal/Mpc$^{2}$. Those  environments could be similar
to those showed  by galaxy groups or weak  clusters of galaxies. Thus,
we  can  say  that  even  for the  densest  environments,  the  global
environment   do   not   play   an   important   role   in   the   bar
formation.  However,  we  can  not exclude  than  close  galaxy-galaxy
encounters would trigger the bar formation, as they were excluded from
our sample.  These results indicate that formation  and evolution of
the  bars in  the  studied  sample depend  mostly  on internal  galaxy
processes rather than external ones.

A  statistical  significant   difference  between  the  central  light
concentration of barred and unbarred galaxies was found.  The bars are
mostly located  in less  concentrated galaxies. This  difference could
explain the lower fraction of bars detected  in S0 galaxies with
respect to  spirals. Since the S0 galaxies host weaker  bars than
spirals, we conclude that  central light concentration is an important
factor driving  the bar formation and evolution.   In fact, according
to   the  numerical   simulations   \citet[][]{pfenniger90,  norman96,
athanassoula02, debattista06}, the bars  are weakened by large bulges.
Finally, bars are mainly hosted  by bluer late-type spirals.  We argue
that this is  due to late-type galaxies have  larger bar fraction than
early-type ones. Similar results were  also found in previous works as
Barazza et al. (2008).

The sample of  galaxies presented in this study is  one of the largest
samples  presented in  the literature.  This large  number  of studied
galaxies makes that the conclusions reported in the present work about
the  different observational  bar  properties is  stronger than  those
obtained with smaller number of galaxies. For this reason, the present
work will  be useful for  constraining future theoretical  works about
formation and evolution of bars in disc galaxies.

\begin{acknowledgements} 

We  thank Victor P.  Debattista, Lorenzo  Morelli, Irina  Marinova and
Shardha Jogee for fruitful discussion. We also thank to the anonymous referee for helpful comments to this manuscript.
J.A.L.A. is funded by the grant AYA2007-67965-C03-01 by the Spanish
Ministerio de Educaci\'on y Ciencia.
J.M.A. acknowledges support from the Istituto Nazionale di Astrofisica
(INAF).
E.M.C. receives support from the grant CPDA068415/06 by Padua University.
J.M.A. and E.M.C. thank the Instituto de Astrof\`\i sica de Canarias
for hospitality while this paper was in progress.
This research has made use of the NASA/IPAC Extragalactic Database
(NED) and Sloan Digital Sky Survey (SDSS).

\end{acknowledgements} 


\end{document}